\documentclass[fleqn, usenatbib]{mnras}

\usepackage{newtxtext,newtxmath}

\usepackage[T1]{fontenc}

\DeclareRobustCommand{\VAN}[3]{#2}
\let\VANthebibliography\thebibliography
\def\thebibliography{\DeclareRobustCommand{\VAN}[3]{##3}\VANthebibliography}

\newcommand*{\CIV}{\ion{C}{iv} }

\usepackage{orcidlink}
\usepackage{graphicx}	
\usepackage{booktabs}
\usepackage{multirow}
\usepackage[utf8]{inputenc}

\title[Radio emission in RQ Quasars]{How does the radio enhancement of broad absorption line quasars relate to colour and accretion rate?}

\author[J. Petley]{
J. W. Petley$^{1}$\thanks{E-mail: james.w.petley@durham.ac.uk}\orcidlink{0000-0002-4496-0754},
L. K. Morabito$^{1}$\orcidlink{0000-0003-0487-6651}, 
A. L. Rankine$^{2}$\orcidlink{0000-0002-2091-1966},
G. T. Richards$^{3}$\orcidlink{0000-0002-1061-1804},
N. L. Thomas$^{1,4}$\orcidlink{0000-0003-4315-4555},
\newauthor
D. M. Alexander$^{1}$\orcidlink{0000-0002-5896-6313},
V. A. Fawcett$^{5}$\orcidlink{0000-0003-1251-532X},
G. Calistro Rivera$^{6}$\orcidlink{0000-0003-0085-6346},
I. Prandoni$^{7}$\orcidlink{0000-0001-9680-7092},
P. N. Best$^{2}$,
S. Kolwa$^{8}$\orcidlink{0000-0001-9821-4987}\\
\\
$^{1}$Centre for Extragalactic Astronomy, Department of Physics, Durham University, Durham, DH1 3LE, UK \\
$^{2}$Institute for Astronomy, University of Edinburgh, Royal Observatory, Blackford Hill, Edinburgh EH9 3HJ, UK \\
$^{3}$Department of Physics, Drexel University, 32 S. 32nd Street, Philadelphia, PA 19104, USA \\
$^{4}$Institute for Computational Cosmology, Department of Physics, Durham University, South Road, Durham, DH1 3LE, UK \\
$^{5}$ School of Mathematics, Statistics and Physics, Newcastle University, Newcastle upon Tyne, NE1 7RU, UK \\
$^{6}$ European Southern Observatory, Karl-Schwarzschild-Str. 2, 85748 Garching bei München, Germany \\
$^{7}$INAF-IRA, Via P. Gobetti 101, 40129 Bologna, Italy \\
$^{8}$Physics Department, University of Johannesburg, 5 Kingsway Ave, Rossmore, Johannesburg, 2092, South Africa\\
}

\date{Accepted XXX. Received YYY; in original form ZZZ}

\pubyear{2023}

\begin{document}
\label{firstpage}
\pagerange{\pageref{firstpage}--\pageref{lastpage}}
\maketitle

\begin{abstract}
The origin of radio emission in different populations of radio-quiet quasars is relatively unknown, but recent work has uncovered various drivers of increased radio-detection fraction. In this work, we pull together three known factors: optical colour ($g-i$), \CIV Distance (a proxy for $L/L_{Edd}$) and whether or not the quasar contains broad absorption lines (BALQSOs) which signify an outflow. We use SDSS DR14 spectra along with the LOFAR Two Metre Sky Survey Data Release 2 and find that each of these properties have an independent effect. BALQSOs are marginally more likely to be radio-detected than non-BALQSOs at similar colours and $L/L_{Edd}$, moderate reddening significantly increases the radio-detection fraction and the radio-detection increases with $L/L_{Edd}$ above a threshold for all populations. We test a widely used simple model for radio wind shock emission and calculate energetic efficiencies that would be required to reproduce the observed radio properties. We discuss interpretations of these results concerning radio-quiet quasars more generally. We suggest that radio emission in BALQSOs is connected to a different physical origin than the general quasar population since they show different radio properties independent of colour and \CIV distance.  
\end{abstract}




\section{Introduction}

Black holes are now key to our understanding of the complex evolution of galaxies, despite being fundamentally simple to characterise \citep[e.g. mass and spin - ][]{Kerr1963GravitationalMetrics}. Nearly all galaxies contain a super-massive black hole (SMBHs) and, during rapid growth, black hole systems can stimulate electromagnetic radiation in their surroundings. These black holes are considered to be "active" and referred to as Active Galactic Nuclei (AGN). Combined with the knowledge that SMBHs grow with their galaxies \citep[see][]{Kormendy2013CoevolutionGalaxies}, AGN have become objects of great interest in many different areas of astronomy today.

Quasars (also known as QSOs) are the brightest of all AGN and are some of the most luminous objects at optical wavelengths in the observable universe \citep{Schmidt1968SpaceSources}. The massive amount of luminosity ($>10^{44}$ erg/s) that quasars radiate could impact the future evolution of their host galaxies since, at least for a time when we observe them, they have a much greater energetic output than all the stars in their galaxy. However, it is often difficult to predict what will be observed across the full spectral energy distribution of a quasar even if some wavelength regions or physical properties (eg. accretion rate and black hole mass) are well understood or measured. Since different wavelength ranges offer information about different physical scales and processes, we cannot build a full picture of quasar behaviour without appreciating the longest, as well as the shortest, wavelengths. 

Radio emission at low frequencies ($\lesssim 10$ GHz) in quasars is largely dominated by the synchrotron radiation process. Emission is released through the acceleration of electrons moving at relativistic speeds through a magnetic field \citep{Condon1992RadioGalaxies.}. Radio emission is largely unaffected by sources of obscuration which significantly modulate what is observed at shorter wavelengths. For example, we avoid the effects of dust that impact optical and ultra-violet (UV) emission. Radio observations could therefore play a key role in understanding energetic interactions between a black hole and its galaxy. The most widely recognisable sources of synchrotron emission are famous \textit{radio-jet} type galaxies with complex jet and filament type structures which extend well beyond the size of the host galaxy \citep{Fanaroff1974TheLuminosity}. The mechanism which extracts energy into these jets from accretion around the black hole is thought to be the Blandford-Znajek process \citep{Blandford1977ElectromagneticHoles.}. The formation of such jets likely has a significant connection with the spin of a black hole \citep{Blandford1990PhysicalNuclei., Wilson1995TheGalaxies}, a property currently not measurable at significant redshift. These jets are increasingly incorporated into both single galaxy or cluster simulations \citep{Husko2022Spin-drivenClusters} and cosmological simulations with AGN feedback \citep{Dave2019SIMBA:Feedback, Thomas2021TheSimulations}.

When studying the radio population, observers have typically made a distinction into two types of source based on the apparent bi-modality in the radio-to-optical luminosity ratio distribution. \textit{Radio-loud} sources have higher radio to optical luminosity ratios than \textit{radio-quiet} sources \citep{Kellermann1989}. Although this can be a useful distinction, often splitting into these two groups can cause confusion amongst astronomers as it has sometimes been implied that radio-loud sources are jets and that radio-quiet sources are not. To be clear, many radio-quiet jetted AGN have been observed \citep{Kukula1998TheQuasars, Leipski2006TheQuasars,Jarvis2019PrevalenceQuasars,Macfarlane2021TheFormation, Girdhar2022QuasarDisc} i.e. \textbf{radio-quiet $\neq$ no jet}. Consequently, while we are confident that the vast majority of radio-loud sources have jets, we lack the required information to make the contrary statement for radio-quiet sources.

The origin of radio emission in a given radio-quiet system is therefore a more complicated matter given there are essentially four emission mechanisms with the potential to be detected by modern radio surveys. These are: jets, star formation, disk winds and corona \citep[see][]{Panessa2019TheNuclei, Kimball2011TheAGN}. Coronal emission is unlikely to be the dominant radio emission mechanism at the low frequencies which we study in this paper \citep{Raginski2016AGNEmission, Behar2018TheAGN, Baldi2022TheData}. Coronal emission is highly compact, optically thick and has a flat spectrum which is not observed for the populations we study. Therefore, we do not include a detailed discussion of this emission mechanism. Distinguishing between the remaining possible components with current large radio surveys is not a simple task since the vast majority of sources are unresolved by most large-scale survey instruments, which means we cannot determine the source of the radio emission directly without higher resolution information \citep{Morabito2022IdentifyingObservations}. 

Radio emission from jets in radio-quiet and radio-loud sources is likely very similar but jets can be frustrated in size and lifetime, by high-density environments of gas and/or dust \citep{White2015Radio-quietMJy, Sadler2016GPS/CSSAGN, CalistroRivera2023UbiquitousWinds}, or simply lower power \citep{Hardcastle2018AGalaxies,Hardcastle2019Radio-loudSources}. Star formation rate correlations with radio luminosity have now been calibrated at several frequencies, including low frequency \citep{CalistroRivera2017The2.5, Gurkan2018LOFAR/H-ATLAS:Relation}, based on the clear correlation between infrared thermal emission and non-thermal radio-emission \citep{Helou1985ThermalGalaxies., Bell2003EstimatingCorrelation}. Although there is a scatter and a strong dependence on the stellar mass of a galaxy \citep{Delvecchio2021The4}, the fact that this correlation exists means that radio is regularly used as a measure for star formation in non-AGN sources. Radio emission from disk winds is probably the most poorly understood of the potential mechanisms. Quasars are known to possess various types or phases of outflows (e.g. Broad Absorption Line Quasars  - \citealt{Weymann1991}, [\ion{O}{iii}] outflows - \citealt{Weedman1970High-VelocityNuclei}, Ultra-Fast Outflows (UFOs) - \citealt{Pounds2003APG1211+143}) and that these outflow phases can also couple to each other \citep[eg. ][]{Feruglio2015TheOutflow, Girdhar2022QuasarDisc}. These high-velocity outflows have the ability to generate shock fronts which could emit synchrotron radiation although currently, only a few simple models of these shocks exist with limited prediction power \citep{Ostriker2010MomentumFeedback,Faucher-Giguere2012TheNuclei, Nims2015ObservationalNuclei}. 

The presence of radio emission has been found to connect to several other quasar properties. The investigation of the interaction between these properties is the focus of this work. A brief introduction to each of these follows.

\subsection{Optical Reddening}

The distribution of rest frame optical colours of quasars is resolved in the Sloan Digital Sky Survey (SDSS) photometry and shows a normal distribution with a longer tail of quasars with excess red colours \citep{Richards2003RedSurvey}. These red colours in this tail of the distribution are most likely caused by the effect of dust \citep{Webster1995EvidenceQuasars, Glikman2007TheSurvey, Krawczyk2015MININGQUASARS, CalistroRivera2021TheReddening,Fawcett2022FundamentalX-shooter}. Quasars that show excess red colours are defined in various ways, usually by their particular percentile of the colour distribution, as \textit{red quasars}. The link between radio emission and red quasars was firmly established by \cite{Klindt2019FundamentalOrientation} who found that red quasars were $\sim3\times$ more likely to be radio detected than blue or \textit{control} quasars using the Faint Images of the Radio Sky at Twenty centimetres \citep[FIRST;][]{Becker1995TheCentimeters}. 

This result has prompted further research to expand to other wavelengths and look at different morphologies and powers. \citet{Klindt2019FundamentalOrientation} and \citet{Fawcett2020FundamentalQuasars} found that the radio enhancement is likely due to emission on compact scales and from the radio-quiet or and radio-intermediate population. \cite{Rosario2020FundamentalLoTSS} used the LOFAR Two Metre Sky Survey Data Release 1 \citep[LoTSS DR1; ][]{Shimwell2019TheRelease} to show that the red quasar radio enhancement exists at low frequencies and at a similar level to \cite{Klindt2019FundamentalOrientation}. High-resolution e-MERLIN data on a small sub-sample of red and blue quasars revealed a statistically significant difference in the incidence of kpc scale emission between the two samples with the red sample having a greater fraction of extended sources at scales $\leq2$ kpc \citep{Rosario2021FundamentalE-MERLIN}. 

The interpretation of these results is that red quasars are inconsistent with a simple orientation model for the dusty torus surrounding the black hole but are broadly consistent as a transition phase where red quasars transition to blue quasars. Some propose that this may fit into larger evolution models such as the one proposed in \cite{Hopkins2008AActivity}. Furthermore, through modelling of the radio-UV spectral energy distribution and optical spectra of quasars, \cite{CalistroRivera2021TheReddening} found that the reddening may originate from dusty polar winds. The increased radio emission could then be due to shocks from AGN winds, or due to shocks from the interactions between compact AGN jets and dusty gas-dense regions \citep{Fawcett2023AQSOs, CalistroRivera2023UbiquitousWinds}, or alternatively increased star formation during this earlier stage of evolution, although the latter has been excluded based on multiwavelength SED fitting by \cite{CalistroRivera2023UbiquitousWinds} and the high-resolution study of \cite{Rosario2021FundamentalE-MERLIN}.

\subsection{Broad Absorption Line Quasars}

Around 10-20\% of quasars selected by optical magnitude show broad absorption features bluewards of the \CIV emission line, although the intrinsic fraction of quasars with these features is expected to be higher \citep[up to 40\%, e.g][]{Dai20072MASSBALQSOs, Allen2010AFraction, Dai2012TheQuasars}. These quasars are known as Broad Absorption Line Quasars (BALQSOs) and have often been characterised using the Balnicity Index (BI) measure of \citet{Weymann1991}. BI is calculated as an integral of a region bluewards of the \CIV emission line when absorption is more than 10\% of the continuum level. Radio studies of BALQSOs have shown that they are more likely to be radio-quiet than non-BALQSOs \citep{Stocke1992TheQSOs, Becker2000PropertiesSurvey} and, similar to the red quasars, they are significantly more likely to be radio-detected \citep{Urrutia2009TheQuasars, Morabito2019TheSurvey, Petley2022ConnectingQuasars}. \citet{Petley2022ConnectingQuasars} found that radio-detected BALQSOs have different absorption profiles compared to non-radio-detected BALQSOs implying an intrinsic connection between the wind and the radio emission, with radio-emission coming from the wind itself (wind shocks) as the most likely explanation for this connection. 

The true fraction of BALQSOs in the quasar population is hard to ascertain. In a purely geometric understanding \citep[e.g. ][]{Weymann1991, Ghosh2007}, the observed BALQSO fraction is equal to the mean fractional solid angle that the BAL wind subtends of the quasar. However, BAL troughs are known to appear and disappear \citep[e.g.][]{Ak2012BroadSample,Mishra2021AppearanceQuasars} and \citep[as noted in][]{Gregg2006FRQuasars} if they are predominantly an evolutionary feature then their lifetime is likely on the order of $\sim$ 10~Myr based on a $\sim 10\%$ fraction of quasars accreting for $\sim 10^8$Yrs \citep{Hopkins2005Luminosity-DependentFunctions}. If indeed the lifetime of a BAL wind is shorter than the lifetime of the average quasar, the BAL wind must have a much larger covering angle than simply the observed fraction in order to maintain that fraction, or alternatively, many quasars go through several BAL phases. However, it is not clear whether this BAL lifetime is a feature of the wind turning on and off or a feature of a particular geometry such as a rotating clumpy wind. Whatever the scenario, their radio-detection enhancement is potentially an indicator that they can interact strongly with their environment even in a relatively short time. In addition, many quasars identified as non-BALQSOs may in fact be hosting BAL winds which are not currently in our line of sight, creating some level of contamination within the non-BALQSO population.

\subsection{\CIV Distance}

The \CIV emission line ($\lambda~1549$) is readily observable in the UV spectra of most quasars. The anti-correlation between the \CIV equivalent width (EW) and UV continuum luminosity, known as the \textit{Baldwin Effect} \citep{Baldwin1977LuminosityObjects}, was evidence that this line could probe some of the accretion properties of the SMBH. Subsequently, the informational content of this line has been explored further and specifically the blueshift of the line was found to be a feature of importance \citep{Gaskell1982AMotions., Wilkes1984StudiesProfiles.,Richards2002BroadQuasars,  Sulentic2003RadioSpace, Sulentic2007Nuclei}. \cite{Richards2011UnificationEmission} investigated radio-quiet and radio-loud quasars in a 2D space of \CIV EW and blueshift and found that the distributions observed could best be understood through a "disk" vs "wind" model with disk dominated systems having high \CIV EW and low blueshifts and wind dominated systems having low \CIV EW and high blueshifts. 

Recent work on this 2D \CIV space has found further evidence of its diagnostic power. \cite{Rankine2020BALSample} found that BALQSOs comprise a higher fraction of moderate EW, high blueshift sources but crucially that they can still be found in all regions of the space. This suggests that BALQSOs are simply QSOs with a probability of observation based on a particular line of sight or at a particular time such that a BAL wind is observed. This probability, and also the average absorption profile of the BALQSOs, changes across the \CIV space. \cite{Richards2021ProbingQuasars} converted the position in 2D space into a single \CIV distance value (low \CIV distance: high EW and low blueshift; high \CIV distance: low EW and high blueshift) and found that the Very Large Array (VLA) radio detection fraction of quasars varied non-linearly with this parameter, hinting at multiple radio emission mechanisms at play, a result also suggested in \cite{Rankine2021PlacingFormation}. Physical drivers of the changing \CIV properties have been suggested. For example, a link between accretion rate and \CIV blueshift was found in \cite{Wang2011CoexistenceNuclei} and \cite{Sulentic2017WhatSequence}.   Most recently \cite{Temple2023TestingQuasars} connected this space to Spectral Energy Distribution (SED) models of accretion and showed that changes in the properties of quasars in this space could be connected to accretion rate and a likely change in the accretion geometry around an Eddington Ratio of $\lambda_{\rm Edd} \approx 0.2$, a transition suggested in \cite{Giustini2019AContext}. We therefore use \CIV distance in this work as a proxy for the accretion rate in our quasars as it is measured for all of our sources and does not have any additional conversions or assumptions.

This work aims to explore each of these quasar properties - \CIV distance, optical colour and BAL winds - in combination and attempt to understand their intrinsic connection to radio emission. Each of these properties show overall changes in detection fraction of a similar magnitude and previous work has not incorporated all of them simultaneously. With an improved picture of the connection between these properties, we can re-think the underlying physical scenarios that govern radio emission in quasars.

\section{Data and Methods}

To create our sample of quasars we begin with the SDSS DR14 quasar catalogue of \cite{Paris2018TheRelease}. This catalogue summarises the properties of 526,356 quasars including redshifts and whether a BAL wind is present, although this is the first of the SDSS releases not to visually inspect all spectra for BAL signatures. To improve the quality of the spectral data and also extract several important parameters for this study we use spectral reconstructions of SDSS DR14 quasars created through mean-field independent component analysis and described by \cite{Rankine2020BALSample}. The process improves the redshift measurements and redefines parameters such as the BI in comparison to the SDSS release. We therefore use these values to define our BALQSO sample. 

Reconstructed spectra are used to obtain the \CIV EWs and blueshifts and to parameterise the \CIV EW-blueshift space (hereafter "\CIV space") with a single value, the \CIV distance. The method for this process is described in \cite{Rivera2020CharacterizingSpectroscopy} and also improved in \cite{Rivera2022ExploringSpace}. Strictly, we use the parallel \CIV distance (\CIV~$\parallel$) described within \cite{Rivera2022ExploringSpace} but simply refer to it as the \CIV distance.  In summary, the distance is formed through a piece-wise polynomial fitting of all quasars in the \CIV space which is then scaled using a \textit{MinMaxScaler} algorithm to place all values between 0 and 1 with an equal weighting to both axes. Code to calculate \CIV distance is available from \cite{McCaffrey2021CIVDistance}.

We use a slightly different technique to characterise red and blue quasars in our sample compared to recent radio studies of red and blue quasars such as \cite{Klindt2019FundamentalOrientation, Fawcett2020FundamentalQuasars, Rosario2020FundamentalLoTSS}. It is not only the most red sources that show radio enhancement and we aim to study the overall effect of changing colour. We attempt to capture the variation in colour of quasars independent of redshift and keep all of the quasars in our working sample. By obtaining the redshift-independent optical colour distribution, we can then assign some quasars as being excess red due to some extra component, most likely dust and the rest as normal variation captured by a Gaussian. This is more similar to the reddening studies of \cite{Richards2003RedSurvey} and \cite{Glikman2022TheRegime}.

To accurately determine the colours of our quasars using SDSS photometry, we restrict our sample to $1.7<z<2.5$. The lower limit of $z = 1.7$ is imposed by the observable range of the \CIV emission and absorption region in all of the SDSS quasars. The upper limit of $z = 2.5$ is imposed by the contamination of the Lyman break in the SDSS $g$ band photometry at higher redshifts. 

\begin{figure}
    \centering
    \includegraphics[width = 0.95\linewidth]{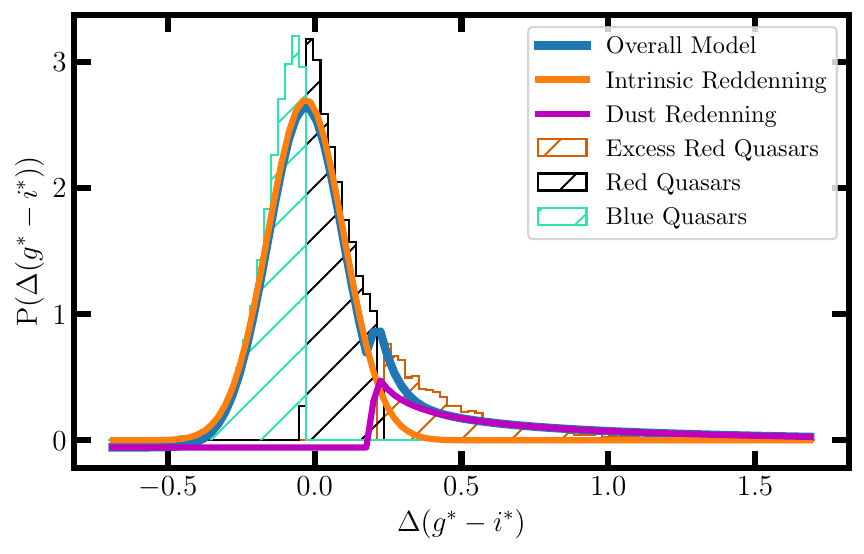}
    \caption{The probability density of redshift corrected quasar optical colour. We define different optical colours by splitting the quasars into redshift bins and then shifting the $g-i$ colour distribution in that bin to a median of zero. We then fit the stacked distribution with a normal (blue) and lognormal curve (pink). Excess red quasars are defined in the region beyond the peak of the lognormal, blue quasars are defined with a shifted colour below zero and red quasars are in between. }
    \label{fig:colour_split}
\end{figure}

We split the sample into 20 equally spaced redshift bins and calculate the galaxy extinction corrected $g-i$ band colour for each quasar in that bin. We then shift the entire distribution within that bin so that the median is at 0. We do this for every bin so that in the end we have a stacked distribution of all the quasars with a median of 0. We then fit two distributions to the data using the \verb|scipy| Python package. First, a normal distribution with a $\mu = 0$ to capture the majority of the colour variation and then another distribution to capture the excess red quasars. We found that a log-normal distribution worked best to capture this tail (see \autoref{fig:colour_split}).

We define all quasars with their colour bluer than the median within their bin as blue quasars (shown in blue in all figures), quasars with their colour more red than the peak of the log-normal distribution as excess red quasars (shown in red in all figures) and those in between these two points as "red" quasars (shown in black in all figures). 

We aim to observe how \CIV distance relates to an independent measure of the AGN luminosity, the  6 micron luminosity ($L_{6 \mu m}$). Therefore we require that our sample has detections in the first three bands of the Wide-field Infrared Survey Explorer (WISE). We perform a log-linear interpolation between the bands to estimate a rest-frame 6 micron luminosity which should capture the re-processed disk emission and be free from other extinction. 

To understand the radio properties of the quasars, we use the LOFAR Two Metre Sky Survey (LoTSS) Data Release 2 described in \citet{Shimwell2022TheRelease}. This is a radio survey of the Northern sky with over 4 million radio sources detected with a resolution of 6 arcseconds with a frequency range centred at 144MHz (wavelength of two metres). Specifically, we use the latest catalogue with optical counterparts identified using a combination of likelihood ratio matching \citep{Kondapally2021TheCatalogues}, machine learning \citep{Alegre2022ATechniques} and citizen scientists \citep{Hardcastle2023TheRelease}. LoTSS operates in a unique sensitivity and frequency range and is the best-suited survey for understanding large radio-quiet populations as it is around an order of magnitude deeper than FIRST \citep{Becker1995TheCentimeters}, an example of an earlier widely used radio survey.

We restrict our quasar spectra sample to the LoTSS DR2 region and cross-match with a radius of 2 arcseconds to the LoTSS DR2 optical ID source catalogue. The overall population split for radio detection and colour is shown in \autoref{table:populations}.   

\begin{table}
\centering
\begin{tabular}{|l|l|l|l|} 
\cline{2-3}
\multicolumn{1}{l|}{} & Radio Detected & Non-radio detected & \multicolumn{1}{l}{} \\ 
\cline{1-3}
BALQSOs & 1463 & 3935 & \multicolumn{1}{l}{} \\ 
\cline{1-1}
Non-BALQSOs & 6565 & 31175 & \multicolumn{1}{l}{} \\ 
\hline
\multicolumn{1}{l|}{} & Blue & Red & Excess Red \\ 
\hline
BALQSOs & 903 & 2389 & 2106 \\ 
\cline{1-1}
Non-BALQSOs & 16959 & 15325 & 5456 \\
\cline{1-1}
\end{tabular}

\caption{We present the population numbers for BALQSOs and non-BALQSOs with respect to their radio detection and colour classification.} 
\label{table:populations}
\end{table}

\section{Results}

\subsection{Populations in \CIV Space}

To understand the behaviour of different populations on their own with respect to \CIV emission, we investigate the distribution of different populations of quasars in the \CIV EW and blueshift space. 

\begin{figure*}
    \centering
    \includegraphics[width = \linewidth]{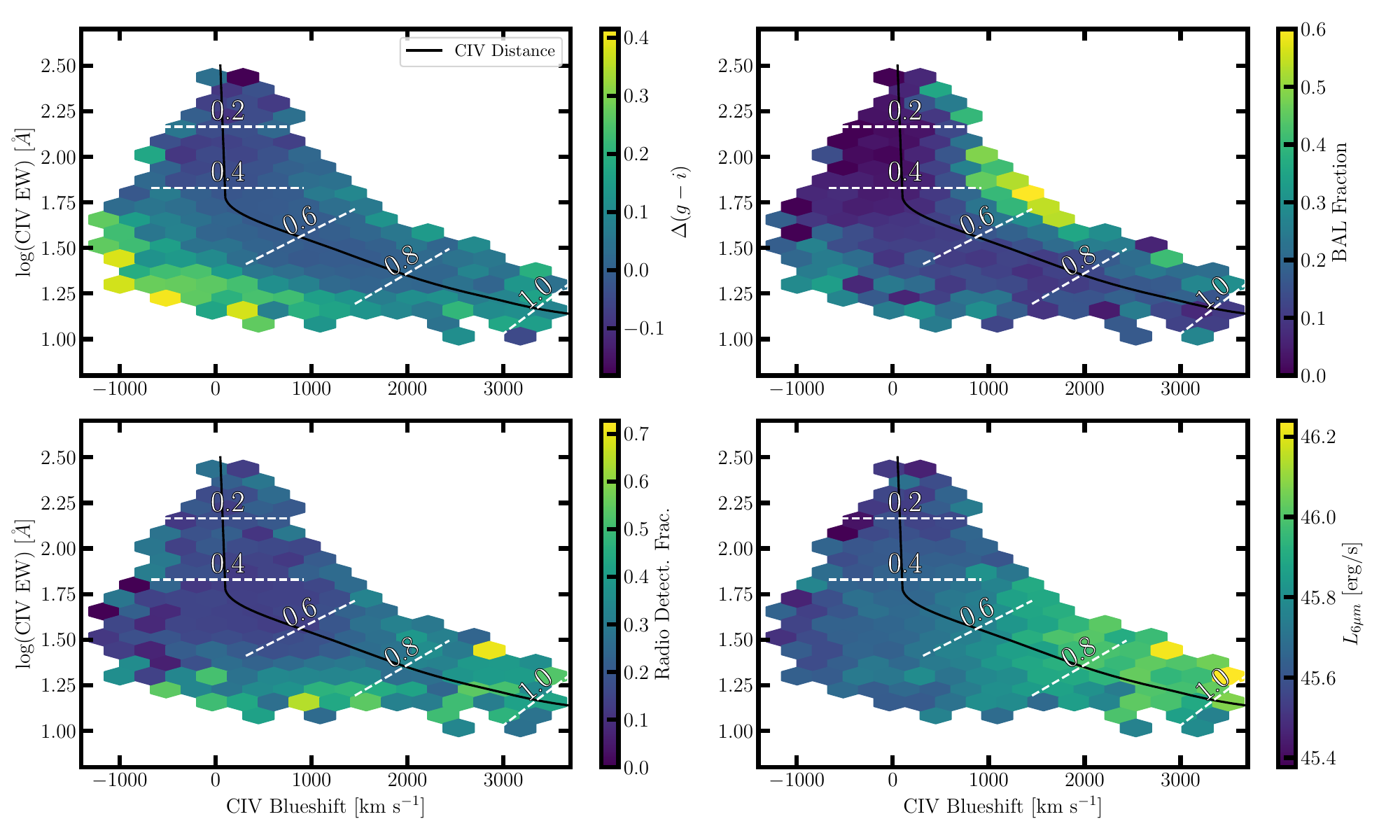}
    \caption{Density plots for different populations in the \CIV EW and blueshift space. The white tracks indicate the value of the \CIV distance at that point. A negative value for \CIV blueshift indicates redshift. In each plot, we require a minimum of 10 sources in each bin. \textit{Upper left:} The median value of $g-i$ colour in each bin. The distribution is relatively uniform apart from the most extreme red sources having low \CIV blueshift and EW. \textit{Upper right:} The fraction of BALQSOs in each bin. This distribution peaks at moderate \CIV EW and higher blueshifts as previously shown in \protect\cite{Rankine2020BALSample}. \textit{Lower left:} The LoTSS radio-detection fraction (RDF) in each bin. This fraction increases with \CIV distance. \textit{Lower right:} The median six-micron luminosity in each bin. This is a measure of the quasar luminosity which increases with \CIV distance.} 
    \label{fig:hexbin_panels}
\end{figure*}

We find that excess red quasars are mostly uniformly spread around the \CIV space (upper left - \autoref{fig:hexbin_panels}) apart from the most extreme red sources being at low \CIV blueshift and \CIV EW. This suggests that the \CIV wind and the origin of the most red colours are not strongly coupled. 

BALQSOs are preferentially found towards higher blueshifts and lower equivalent widths (upper right - \autoref{fig:hexbin_panels}), although the densest BALQSO region in \CIV space is at a moderate value for both. This result is not unexpected as the distribution of BALQSOs in \CIV space was studied extensively in \cite{Rankine2020BALSample}. As well as the BALQSO fraction they also look at the strength and velocity of the BALQSO winds and find that both increase at higher \CIV distance. Even though the BAL fraction change may not be large across the whole space, the average properties of the BAL winds do change quite significantly. There is also a tentative increase in BAL prevalence at low EW and blueshift, potentially connecting to the more red population.

The radio-detection fraction in \CIV space was also studied in \cite{Rankine2021PlacingFormation}, but with the smaller LoTSS DR1. We find the same result here with a much larger sample (lower left - \autoref{fig:hexbin_panels}). Radio-detection fraction increases with \CIV distance.

We investigate an alternative measure of the accretion power of the black hole, the $6\mu$m luminosity (lower right - \autoref{fig:hexbin_panels}). This should probe the warm dust heated by the quasar and provide a measure of the overall AGN luminosity, $L_{\mathrm{AGN}}$ \citep{Richards2006, Gallagher2007AnQuasars, Klindt2019FundamentalOrientation,CalistroRivera2021TheReddening}. We find that it correlates with the \CIV distance quite well. If the black hole mass is constant along the space then it would track the accretion rate. In general, we favour the use of \CIV distance as a measure of the accretion state as the line is driven closer to the black hole and is less contaminated by other infrared radiation.

\subsection{Radio Properties and \CIV Distance}

\begin{table*}
    \begin{tabular}{@{}lccccccccc@{}}
    \toprule
    \multirow{2}{*}{Quasar Pop.} &
      \multicolumn{3}{c}{0<\CIV Dist.<0.2} &
      \multicolumn{3}{c}{0.2<\CIV Dist.<0.4} &
      \multicolumn{3}{c}{0.4<\CIV Dist.<0.5} \\
      \cmidrule(lr){2-4} \cmidrule(lr){5-7}\cmidrule(lr){8-10}
      & {B} & {R} & {ER} & {B} & {R} & {ER} & {B} & {R} & {ER} \\
      \midrule
    BALQSOs & 26 \textbf{/ 6} & 11 \textbf{/ 4} & 7 \textbf{/ 6} & 99 \textbf{/ 23} & 210 \textbf{/ 72} & 105 \textbf{/ 73} & 149 \textbf{/ 16} & 346 \textbf{/ 81} & 227 \textbf{/ 79} \\
    Non-BALQSOs & 242 \textbf{/ 45} & 132 \textbf{/ 30} & 39 \textbf{/ 20} & 3028 \textbf{/ 397} & 2166 \textbf{/ 470} & 538 \textbf{/ 236} & 3423 \textbf{/ 352} & 3605 \textbf{/ 713} & 880 \textbf{/ 317}\\
    \bottomrule

    \toprule
    \multirow{2}{*}{} &
      \multicolumn{3}{c}{0.5<\CIV Dist.<0.6} &
      \multicolumn{3}{c}{0.6<\CIV Dist.<0.8} &
      \multicolumn{3}{c}{0.8<\CIV Dist.<1} \\
      \cmidrule(lr){2-4} \cmidrule(lr){5-7}\cmidrule(lr){8-10}
      & {B} & {R} & {ER} & {B} & {R} & {ER} & {B} & {R} & {ER} \\
      \midrule
    BALQSOs & 272 \textbf{/ 35} & 835 \textbf{/ 201} & 536 \textbf{/ 228} & 190 \textbf{/ 64} & 581 \textbf{/ 250} & 417 \textbf{/ 268} & 13 \textbf{/ 9} & 45 \textbf{/ 55} & 57 \textbf{/ 80} \\
    Non-BALQSOs & 4612 \textbf{/ 475} & 5183 \textbf{/ 1078} & 1156 \textbf{/ 486} & 3363 \textbf{/ 662} & 4127 \textbf{/ 1342} & 881 \textbf{/ 539} & 235 \textbf{/ 115} & 444 \textbf{/ 285} & 157 \textbf{/ 129}\\
    \bottomrule
  \end{tabular}

\caption{We present the populations in each \CIV distance bin for BALQSOs and non-BALQSOs. Radio non-detections are in normal font, radio detections are in bold and colour classifications are: B - blue, R - red, ER - excess red.} 
\label{table:bins}
\end{table*}

We hypothesise that different processes dominate radio emission at different ends of the \CIV distance space with low \CIV distance objects having a greater jet component and high \CIV distance having a greater wind component. We therefore split into \CIV distance bins and measure the radio-detection fraction for different populations. The bins were selected to ensure a similar number in each bin in the moderate \CIV distance range where most sources reside and to have reasonable statistics at the low and high \CIV distance end. The populations within our chosen bins are listed in \autoref{table:bins} along with the bin definitions.

\begin{figure*}
    \centering
    \includegraphics[width = 0.95\linewidth]{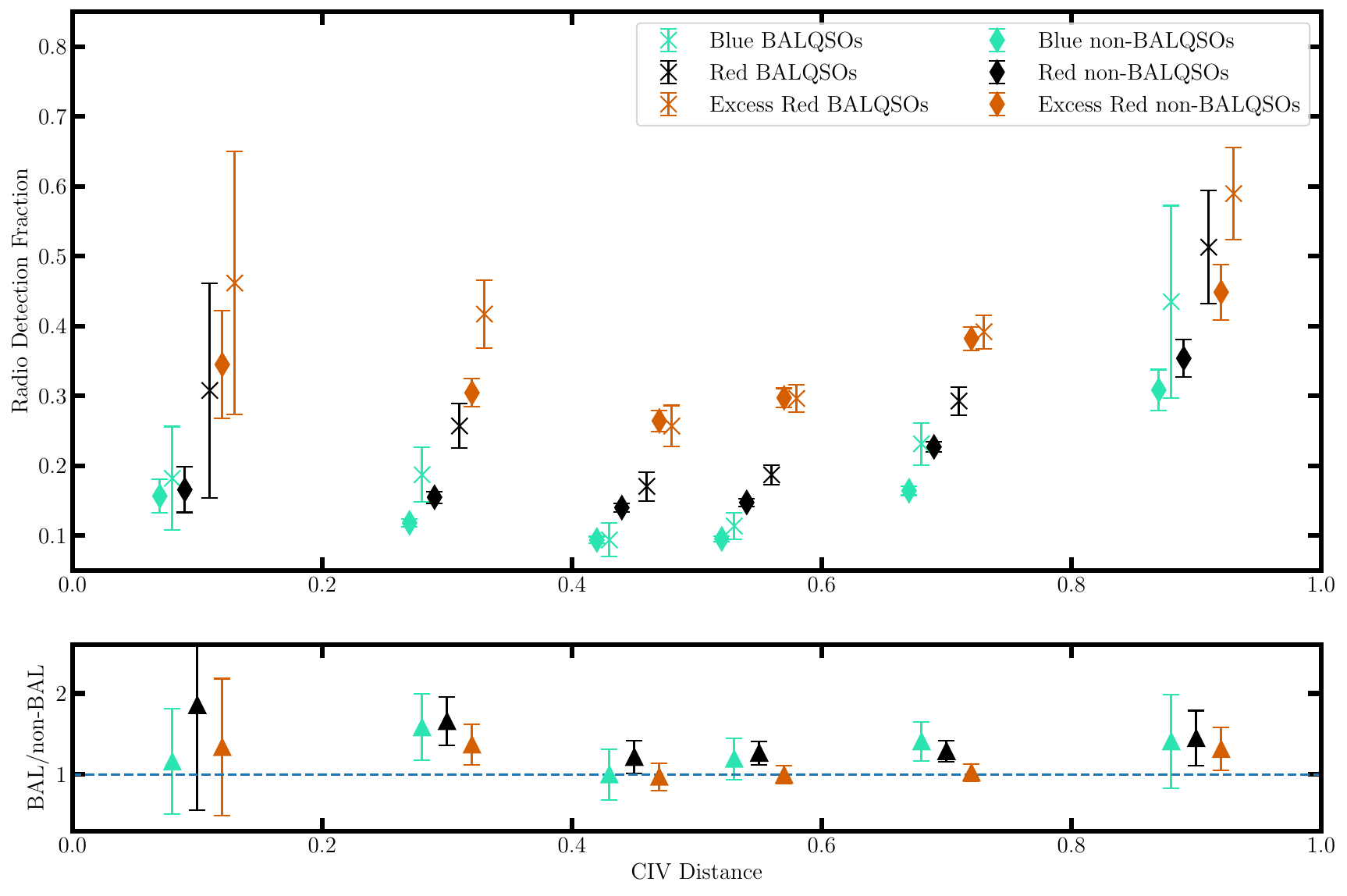}
    \caption{Radio-detection fraction for six \CIV distance bins split into populations based on colour and whether a BAL wind is present. The spread around each \CIV distance value is just to aid the eye. The colour represents the colour of the quasar: blue for blue, black for red and red for excess red quasars. Non-BALQSOs are represented by diamonds and BALQSOs are represented by crosses. In the lower panel, the ratio of radio-detection fraction for BALs and non-BALs is displayed as a clearer quantification of the BAL effect in each bin.}
    \label{fig:detection_fraction}
\end{figure*}

In \autoref{fig:detection_fraction} we show these radio-detection fractions. The error bars in these figures indicate the standard deviation within each bin. We find that all 3 of the parameters, \CIV distance, colour and the presence of a BAL wind, have their own effects as expected but we also find different behaviour at each end of the \CIV emission space. Radio detection fraction increases from blue to excess red quasars. The change in radio-detection fraction from blue to red is of a similar magnitude, around 5\%, to that of the change from red to excess red quasars. This is consistent with the recent work of \cite{Fawcett2023AQSOs} which analysed the connection between dust extinction and radio emission using quasars identified with the Dark Energy Spectroscopic Instrument \citep[DESI:][]{DESICollaboration2016TheDesign} and LoTSS, finding a strong positive correlation between the two properties.

Overall the radio detection fractions decrease with \CIV distance to 0.5 and then increase with increasing \CIV distances, with the largest change, both in overall percentage and percentage per unit \CIV distance (the gradient), coming between the two highest bins. This behaviour is similar for all populations, suggesting that different accretion properties are not the driving force behind the radio differences between BALQSOs and non-BALQSOs or behind reddening in quasars.

\begin{figure*}
    \centering
    \includegraphics[width = 0.95\linewidth]{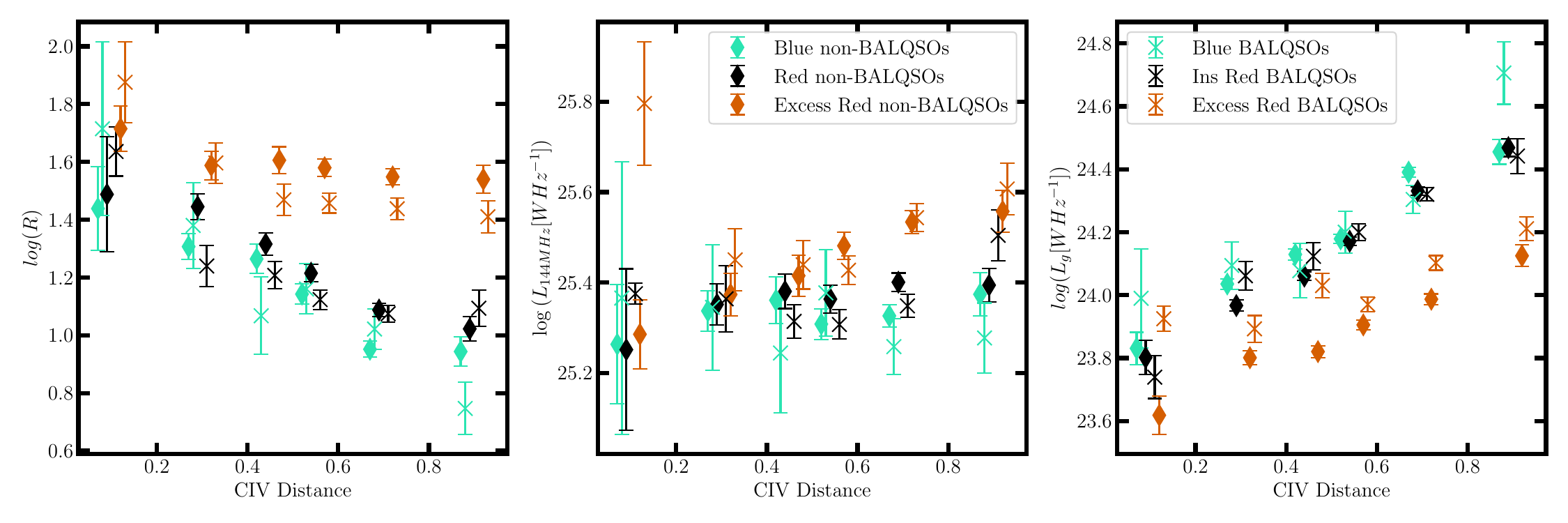}
    \caption{In all plots non-BALQSOs are represented with diamonds and BALQSOs are represented with crosses. \textit{Left:} The radio loudness as a function of \CIV distance shows a decrease for all populations. We suggest that the trend observed is largely due to the changing SED of the quasar as the accretion rate increases.  The classical radio-loud threshold would be $\log(R) > 2.5$. \textit{Centre:} The median radio luminosity as a function of \CIV distance, assuming a spectral index of $\alpha = -0.7$, of different quasar populations in \CIV distance bins. We find that radio luminosity remains constant at moderate \CIV distance which accounts for the majority of most quasars. We only see a definitive increase in the highest bin at the highest accretion rates.  \textit{Right:} $g$-band optical luminosity as a function of \CIV distance for radio-detected quasars. The optical luminosity is increasing, driving down the radio-loudness suggesting that radio-loudness is not a useful measure in determining radio-emission mechanisms without controlling for accretion rate. }
    \label{fig:radio_properties}
\end{figure*}

In \autoref{fig:radio_properties} we also look at the median radio luminosity, median $g$-band optical luminosity and the radio loudness parameter (the ratio of radio to optical luminosity) as a function of \CIV distance for sources that are radio detected in the same bins as \autoref{fig:detection_fraction} and for the same populations. We find again that the largest changes for the whole population of quasars occur at the low and high \CIV distance extremes although it should be noted that the uncertainty is much more considerable at low \CIV distance.  

Firstly, it should be noted that the overall range in median radio luminosity for the whole sample is less than one dex. The radio luminosity results generally indicate that, when detected, BALQSOs have similar radio luminosities compared to non-BALQSOs despite their higher radio detection fractions. For non-BALQSOs, the increase in radio luminosity from blue to red and from red to excess red quasars is similar and only $\sim0.05$ dex.

The radio loudness parameter is typically used to identify excess radio emission in sources relative to their optical emission. In quasars, the optical luminosity is thought to be a good measure of the power of the quasar since it is assumed we are obtaining a largely unobscured view of the accretion disk. In \autoref{fig:radio_properties} we use the $g$-band luminosity to calculate the radio-optical radio-loudness. The classical divide from radio-quiet to radio-loud populations would occur at $\log(R) > 2.5$ at the LoTSS 144MHz frequency. Clearly, our sample is dominated by radio-quiet sources.

We observe a decrease in the radio loudness parameter as \CIV distance increases. This was similarly observed when only looking at \CIV blueshift in \cite{Rankine2021PlacingFormation}. However, we show that this change in radio loudness is largely driven by an increase in the $g$-band optical luminosity. The radio luminosity for all populations across \CIV distance changes by less than 1 dex. Therefore it is the increase in optical luminosity with \CIV distance that is driving this behaviour. This shows some of the limits of radio-loudness as an idealistic measure of "jettedness" as two sources with similar radio power, accretion rate and radio emission mechanism can have different radio-loudness measures based on different optical continuum spectral slopes \citep{Balokovic2012DisclosingSample}.

\section{Discussion}

We consider the following different interpretations of these results with regard to the main potential radio emission mechanisms in radio-quiet quasars: star formation, jets and wind shocks. We provide a brief introduction to each mechanism here but for a detailed review see \cite{Kimball2011TheAGN,Panessa2019TheNuclei}.

Synchrotron emission from star-forming regions occurs due to the presence of free electrons and magnetic fields in supernova remnants. The radio flux from star-forming galaxies has been found to correlate extremely well with the far infrared emission, a presumed indicator of the star formation rate \citep{Sopp1991AQuasars.,Appleton2004Survey, Sargent2010NoField, CalistroRivera2017The2.5,Smith2021TheFrequencies}. This correlation has been shown to hold for low radio luminosity quasars to moderate redshifts \citep{Bonzini2015StarSources, Gurkan2018LOFAR/H-ATLAS:Relation}. However, many studies find radio-quiet AGN with a compact core that lie on this correlation \citep{Maini2016CompactAGNs, HerreraRuiz2016UnveilingQuasars, McCaffrey2022Kpc-scaleQSOs}, suggesting an AGN origin for at least some of the radio-quiet emission. A multi-resolution radio study of sources can allow for a better characterisation of star formation and AGN components \citep{Morabito2022IdentifyingObservations}.

Radio jets dominate the radio-loud population, but whether their contribution extends to many of the radio-quiet sources is a topic of much research. It is possible to model the radio emission and attempt to split contributions from star formation and AGN components \citep[eg.][]{Mancuso2017GalaxyNuclei}. Specifically for LOFAR, \cite{Macfarlane2021TheFormation} fit a two-component, star formation and jets, model to the radio-quiet LoTSS population. They find that jets could contribute the majority of radio emission in 10-20\% of all quasars across redshift and optical luminosity. Yue et al. (submitted) extend this analysis through a Bayesian framework and apply their models to quasar reddening, showing that the radio enhancement of reddened quasars is due to the AGN component and not star formation. This is also shown by \cite{Fawcett2020FundamentalQuasars} and \cite{Rosario2020FundamentalLoTSS} for radio-faint quasars. High-resolution studies of radio-quiet quasars have uncovered systems with compact core regions implying the presence of small-scale jets \citep{Kukula1998TheQuasars,Leipski2006TheQuasars,HerreraRuiz2016UnveilingQuasars,Jarvis2019PrevalenceQuasars}. 

AGN outflows have been observed at many different scales and in different phases. When a wind interacts with the intragalactic medium, if it has sufficient velocity it will generate a shock. These shocks can generate relativistic electrons and corresponding synchrotron emission. If the wind is generated from the accretion disk, then we expect an anti-correlation between the jet and wind prevalence since jets are typically formed at lower accretion rates when the disk truncates at a greater distance from the black hole and can generate larger magnetic fields. This is indeed observed in \cite{Mehdipour2019RelationAGN} who found an inverse correlation between radio loudness and the column density of the ionising wind in radio-loud AGN.

With this wind-jet inverse correlation in mind, the hypothesis for radio emission mechanisms across \CIV distance space is that jets dominate the radio emission at low \CIV distance (low Eddington ratio) and that winds contribute more and more as the \CIV distance increases (higher Eddington ratios).

\subsection{Does radio emission in radio-quiet quasars connect to accretion?}\label{sec:accretion}

From \autoref{fig:detection_fraction} we observe all populations to increase in radio-detection fraction with increasing \CIV distance above a \CIV distance of 0.5 and decreasing with increasing \CIV distance below that point. Therefore, it is clear that in some way radio emission across the whole quasar radio-quiet population responds to the accretion properties of the quasar. Even though the changes may not be consistent for each group at all points in \CIV distance, the fact that they all do couple to accretion is illuminating.  For example, if one is to assume that star formation is the dominant factor in the radio-emission of radio-quiet quasars, then an increase in radio-detection fraction for quasars with potentially high accretion rates implies that the same gas is fuelling both star formation and accretion or that a significant positive feedback effect can occur on the timescale at which a quasar exists with a high \CIV distance. 

In general, the radio luminosity changes are minimal across moderate \CIV distance from 0.3-0.6. This is the region in which the system is neither jet nor wind-dominated. \cite{Richards2021ProbingQuasars} also found this was the region of lowest radio detection fraction for VLA quasars. Since we are tracing the behaviour of radio detection fraction with a proxy for accretion rate, could this be the area where AGN contributions are minimal and the radio emission is dominated by star formation? Under that assumption, we would conclude that BALQSOs and red/excess red quasars in this range have higher star formation rates (SFRs) than blue non-BALQSOs. The SFR-radio luminosity correlation for LOFAR is studied in \cite{Smith2021TheFrequencies} and is roughly linear in the radio luminosity region our sample covers. The fact that when they are radio-detected, BALQSOs and red/excess red quasars do not have significantly different radio luminosities to the rest of the population at moderate \CIV distances, \autoref{fig:radio_properties}, suggests that SFR alone cannot explain the increase for both BALQSOs and red/excess red quasars. 

Increased SFRs for red quasars and BALQSOs is a key prediction of the hypothesis that both these populations are linked to earlier stages in the evolution of galaxies than typical quasars if the timescale of the red phase is long enough to observe a decrease in SFR. However, several studies have tried to test this connection and have not found any link between red quasars and enhanced SFRs \citep{Fawcett2020FundamentalQuasars, CalistroRivera2021TheReddening, Andonie2022ASystems}. It could also be that star formation contributions are similar to AGN in this region (moderate \CIV distance), emphasising differences in the connection between black holes and radio emission between the various populations. We hope to explore this further in the future through extensive SED fitting of sources in the LoTSS Deep-Fields \citep{Tasse2020TheImaging, Best2023TheProperties}.

It has been noted in previous work that the radio-detection fraction could have a significant correlation with the luminosity of the AGN itself. To investigate the effect that this has, previous red quasar studies have used a redshift and six-micron luminosity-matched sample to study the radio properties independently of the AGN luminosity \citep{Klindt2019FundamentalOrientation, Fawcett2020FundamentalQuasars}. When performing any luminosity matching we have to be cautious as we are explicitly looking to test correlations with the \CIV distance which we are claiming is a proxy for the accretion rate and will therefore correlate strongly with the six-micron luminosity. Indeed it is clear from \autoref{fig:hexbin_panels} that this correlation does exist and trying to create a matched sample across six different populations (BALQSOs vs non-BALQSO $\times$ three colour bins) would remove a lot of sources and some of these trends we are interested in seeing. 

We checked the luminosity and redshift distributions within each \CIV distance bin of \autoref{fig:detection_fraction} and did indeed find some differences between populations. We verify our key result by performing a 2d-rejection sampling method to match all populations to the redshift-six-micron distribution of the red quasars. We then repeat our previous analyses.

\begin{figure}
    \centering
    \includegraphics[width= 0.95\linewidth]{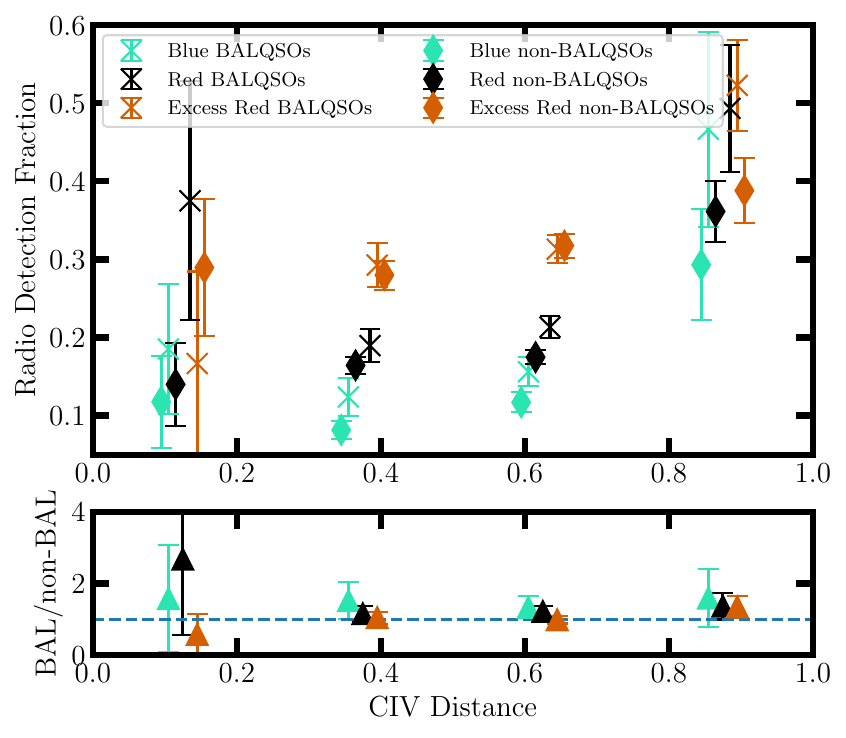}
    \caption{The radio detection fraction with \CIV distance for samples matched in redshift and six-micron luminosity space. We use four bins of 0-0.25, 0.25-0.5, 0.5-0.75 and 0.75-1 in \CIV distance. The radio enhancement of red quasars is still apparent and also the moderate enhancement of BALQSOs remains. The enhancement of low \CIV distance samples appears to be somewhat diminished. Below is a panel displaying the ratio of radio detection fraction between BALs and non-BALs. }
    \label{fig:matched_detections}
\end{figure}

The results from the matched sample are presented in \autoref{fig:matched_detections} which is an equivalent plot to \autoref{fig:detection_fraction} but with fewer bins to maintain similar uncertainties. The same key features are present with this matching. The increase in radio-detection fraction from blue to excess red quasars and moderate enhancement of BALQSOs remains, demonstrating the robustness of our results to luminosity effects.

\subsection{Are BALQSOs a distinct class of quasars?} 

\cite{Rankine2020BALSample} studied in detail the 2D \CIV space properties of BALQSOs and non-BALQSOs and concluded that, since you can find a non-BALQSO with the same \CIV position and SED hardness as any BALQSO, they must both originate from the same parent sample. \cite{Temple2021ExploringQuasars} found that the hot dust emission is also similar between BALQSOs and non-BALQSOs at the same \CIV position. This supports a geometric interpretation for the BALQSO phenomena at a given position in \CIV space. However, the recent radio studies of BALQSOs suggested that the more we explore the properties of BALQSOs at longer wavelengths the more they appear to differentiate themselves from non-BALQSOs. \cite{Morabito2019TheSurvey} highlighted the increasing radio-detection fraction with BI, well over 50\% of quasars with the strongest winds being detected with LOFAR. In \cite{Petley2022ConnectingQuasars} we showed further that this radio emission connects to the absorption profiles of the BALQSO and that this radio emission is extremely difficult to explain with an angular interpretation of BALQSOs with a standard equatorial wind. How then can we reconcile this apparent contrast between the UV and radio nature of BALQSOs?

In this work, we show that much of the BALQSO radio enhancement found in previous papers can be explained by their connection to optical colour. The enhancement is still present in the radio-detection fraction of BALQSOs compared to non-BALQSOs of the same colour classification and contributes around 5-10\% in each bin. The effect is more easily seen as a ratio between BALQSO and non-BALQSO detection fraction in \autoref{fig:detection_fraction}. The median ratio across the bins is 1.29, 1.46 and 1.17 for blue, red and excess red quasars respectively. The ratios are a much smaller enhancement than that observed when not accounting for optical colour such as the detection fractions presented in \cite{Morabito2019TheSurvey} and \cite{Petley2022ConnectingQuasars}. The difference in these studies opens many further questions about the connection between these two samples as in the reddest colour bins the BALQSO fraction can be well over 25\%. This feature has been observed in other samples such as in \cite{Vejlgaard2024AbsenceQuasars}. They create a red quasar sample that would not be selected in SDSS and find that it does not have a higher radio-detection fraction than a comparative SDSS blue sample using FIRST. The red sample had a much higher BAL fraction than in SDSS and they suggest this may be the reason that they do not observe a radio difference. To test this we looked at the BALQSO and non-BALQSO radio-detection fraction relationship with colour alone. 

\begin{figure}
    \centering
    \includegraphics[width = \linewidth]{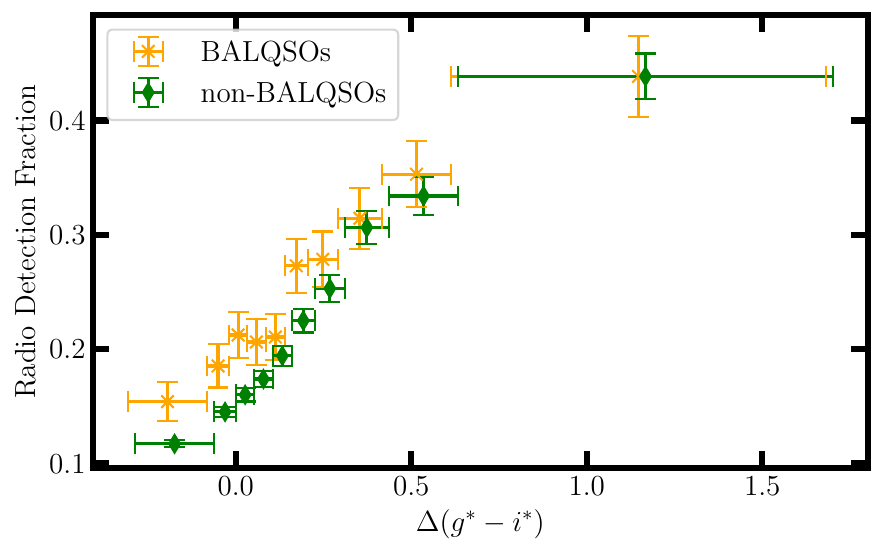}
    \caption{Radio-detection fraction against corrected colour for BALQSOs (orange) and non-BALQSOs (green). Each bin was defined by having an equal number of BALQSOs and the points are separated slightly horizontally to aid the eye. The BALQSO contribution to the radio-detection fraction is most prominent for bluer colours.}
    \label{fig:radio-with-colour}
\end{figure}

In \autoref{fig:radio-with-colour} one can see that the BALQSO radio-detection contribution within SDSS occurs primarily within the normal distribution of quasar colours and not in the most excess red systems. This figure is somewhat similar to those found in \cite{Fawcett2023AQSOs} and \cite{CalistroRivera2023UbiquitousWinds} but with a focus on the BALQSO effect as colour increases. We find a similar behaviour of increasing detection-fraction with colour for both populations. We, therefore, conclude that the BALQSOs fraction is unlikely to be the origin of the difference in radio emission for red and blue populations of quasars as in \cite{Vejlgaard2024AbsenceQuasars}. The differences could reside in the use of LoTSS which is deeper than FIRST and also the selection of the two samples. There are some non-optically spectroscopic surveys (WEAVE-LOFAR - radio-selected \citealt{Smith2016TheSurvey}; 4G-PAQS - astrometrically-selected \citealt{Krogager2023The4G-PAQS}) upcoming which may provide samples with much higher BALQSO fractions than SDSS. These will allow us to study the BALQSO-radio connection in even greater detail.
 
If one considers the idea that increasing \CIV distance relates to increasingly wind-dominated systems, one could interpret the radio-detection fraction of blue non-BALQSOs at a low or moderate \CIV distance to be a measure of the fraction from the combined effect of small-scale jets and star formation. If we then make the assumption that the contribution to the radio from star formation and jets does not increase with \CIV distance, then the increasing radio detection fraction with increasing \CIV distance could be related to shocks. With regards to jets, the motivation that their contribution does not increase with \CIV distance is supported by jets being largely stochastic sources and also radio loudness having an anti-correlation with accretion rate \citep{Sikora2007RadioImplications}. For star formation, our assumption that it has no correlation with \CIV distance finds some support from \cite{Stanley2015AAGN} which found that the relationship between star formation rate and AGN luminosity was remarkably flat. They argue that the AGN luminosity varies much more rapidly than the star formation rate, washing out any underlying correlations. If a BALQSO wind is observed then it means conditions around the SMBH are suitable for the creation of the BALQSO wind. Are these conditions stable for a sufficient period of time for us to both observe the BAL wind and any resultant radio emission (see \autoref{sec:timescales})? Presumably, BALQSO winds transport gas and dust from a more central region out into the ISM adding another uncertain timescale to pair with any proposed connection between BALQSOs, reddening and radio emission. 

In \cite{Choi2022TheSimBAL} a large sample of BALQSOs with lower ionisation and iron absorption troughs, FeLoBALs, was studied using the spectral analysis code SimBAL \citep{Leighly2018}. This represents the most detailed analysis of the physics of BALQSO outflows although admittedly for a highly extreme sub-sample. They found that nearly 18\% of the FeLoBALs contained multiple BAL components and that the objects with the highest kinetic energy fractions were found closer to the SMBH in the vicinity of the torus. Dust could have a key part to play in the expansion of a wind \citep{Everett2010DustyBubbles} and a recent study of the interaction between a torus-like dusty structure and quasar radiation found that dust is the key to generating high-velocity winds without over-ionising the gas \citep{Soliman2023DustVariability}. This could be a key connection between BALQSOs and their increased fraction in reddened samples. Since BALQSOs likely require some amount of shielding, and LoBALs even more so, dust could be a key factor in launching the BALQSOs winds in the first place. 

\subsection{Importance of Colour}

The fact that in \autoref{fig:detection_fraction} and \autoref{fig:matched_detections} we see clear changes in the radio detection fraction even for the red population, suggests that colour, which is likely connected to dust, couples very strongly to radio emission. \cite{Klindt2019FundamentalOrientation} showed the radio-detection fraction for SDSS quasars as a function of $g-i$ colour for different redshift cuts. They found the 1.4GHz FIRST detection fraction increases steeply around the 80th percentile. We use a different definition for our red quasar population in this study, which means our red quasars are not as red as in that study, but we do find that even the quasars with these intermediate colours have a clear increase in radio-detection fraction. FIRST is a less sensitive survey than LoTSS and operates at a higher frequency. This makes it harder to observe detection fraction differences between blue and moderately reddened systems since FIRST observes a greater radio-loud fraction thereby potentially missing lower luminosity emission. 

\begin{figure}
    \centering
    \includegraphics[width = \linewidth]{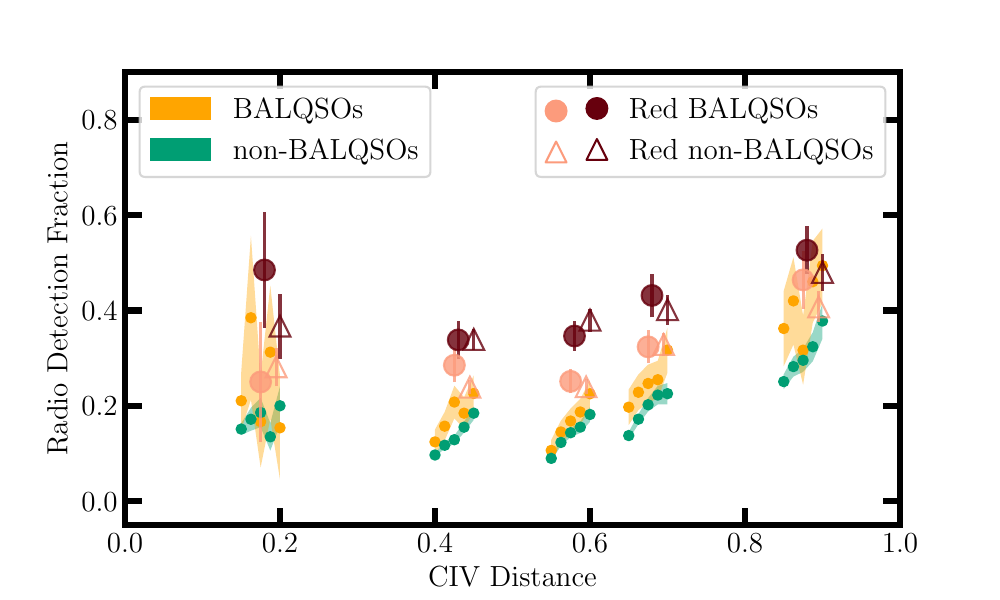}
    \caption{Radio-detection fraction for a smaller number of \CIV distance bins with different colour splits for blue/red quasars and excess red quasars. We use 5 colour bins with an equal number of quasars in each bin for the blue and red groups. From left to right within each shaded region at each \CIV bin the colour is increasing. The BALQSOs are in orange and the quasars are in green with the shaded region representing $1\sigma$ uncertainties. We use 2 equal number bins for the excess red quasars with circles indicating BALQSOs and triangles for quasars. The darker shade is the most reddened bin. Colour increases enhance the radio detection fraction even at the blue end and the difference between BALQSOs and red quasars is minimal in the most red colour bins.}
    \label{fig:colour_bins}
\end{figure}

To make use of the high number of sources at moderate \CIV distance and the fact that we resolve the colour distribution with SDSS photometry, we look again at the radio-detection fraction of quasars but with a focus on the colour in \autoref{fig:colour_bins}. This is similar to \autoref{fig:detection_fraction} but this time we take the excess red population of quasars and split it into two colour bins of equal number and then for the rest of the population, the "blue" and "red" quasars within the normal distribution of \autoref{fig:colour_split}, we split into five colour bins of equal number. We use fewer \CIV distance bins to keep the uncertainties lower and focus on the central region which contains the vast majority of quasars.

The reddened quasars show an enhanced radio-detection fraction at all \CIV distances. This is also seen when comparing blue to red BALQSOs. Hence, dust content could be important for radio emission even when a system is not overwhelmingly disk-dominated and producing high \CIV blueshifts. In fact, the change in the ratio between the red quasar detection fraction and the blue quasar detection fraction is relatively constant across \CIV distance. One hypothesis for red quasars is that they exist as a brief intermediate phase between a dusty star-forming source and a blue quasar \citep{Sanders1988WarmQuasar, Hopkins2008AActivity}. It has recently been shown that the more extreme counterparts of red quasars, known as obscured quasars, have potentially 3 times the star formation rate at the same mass as unobscured quasars and also enhanced radio emission \citep{Andonie2022ASystems}. However, the work of \cite{Rosario2021FundamentalE-MERLIN} showed that the physical scale on which the majority of the radio enhancement of red quasars exists is on a scale $\leq 2$ kpc implying an AGN origin for a significant fraction of the radio emission in red quasars. An alternative explanation could be that the radio emission comes from increased circumnuclear star formation rates and that as the \CIV distance increases we are heightening the effect of positive feedback from the winds that trigger nuclear starbursts \citep[e.g.][]{Silk2003AGone}. But, as discussed in \autoref{sec:accretion}, this appears to be unlikely, at least for the non-BALQSO colour connection to radio emission.

\subsection{Timescale Issues}\label{sec:timescales}

One aspect of radio emission that we have not discussed to this point is the timescale over which radio-emitting sources need to be powered to reach luminosities that we can observe. Radio jets originating from the SMBH typically reach peak luminosities around 10-100Myrs after they are first powered \citep{Hardcastle2018AGalaxies}. If the power stops then the synchrotron luminosity is expected to decrease rapidly. In the models of  \cite{Hardcastle2018AGalaxies} the luminosity drops around 3 orders of magnitude in around 10Myrs after the jet stops being powered (see Figure 6 of that paper). The light travel time for a jet of size 100kpc is 326,000 years which is much lower than this decrease. Since these are radio-quiet AGN and are largely unresolved at 6 arcseconds, we know that they cannot have powerful jets that have been active for 100s of Myrs.

The lifetime of BAL winds themselves is also largely uncertain. The most recent estimates are that BAL troughs exist for around 100 years but that a BAL phase lifetime is much harder to estimate as BAL troughs could also appear during that time frame \citep{DeCicco2018CSample, Mishra2021AppearanceQuasars}. This variability is generally thought to arise due to changes in the gas shielding the BAL from the quasar and hence changing the ionising flux incident upon the wind. The scale on which the overall BAL wind varies or whether winds are always present and just the ionisation state changes is uncertain.

The timescale over which a wind shock can generate significant luminosity is expected to be much shorter than the time scale of a radio jet. Supernova shocks create radio emission and this has been observed and modelled for decades. Typically supernova shock radio emission quickly reaches a peak ($\sim$0.1 years) before decaying more slowly \citep{Chevalier1982TheSupernovae.,Weiler1986RadioSupernovae}. Presumably this time is much longer due to the larger size of an AGN outflow but perhaps this is a way to constrain the connection between a BALQSO lifetime and its radio emission. If we can definitively show that the radio emission is triggered by the BALQSO wind itself then this would allow us to connect the lifetime of the BALQSO to the lifetime of the radio emission by looking at the ratio of radio-detected and non-detected BALQSOs. The radio-detection fraction for a population is modulated by the overlap in the lifetime of that population and the lifetime at which the radio emission is luminous enough to be detected. The likely short lifetime of a BALQSO compared to a quasar in general therefore also implies that the radio emission exists over a similarly short timescale so that the difference in radio detection fraction between BALQSOs and non-BALQSOs can be observed.

\subsection{Testing a wind shock emission model}

\cite{Nims2015ObservationalNuclei} provide a model for wind shocks from AGN winds and has been widely cited in the field. The model makes some simple assumptions for an energy-conserving wind. They first show that for a wind travelling at 10\% the speed of light through a typical ambient medium over a quasar lifetime, the wind can achieve $>1$~kpc scales. They then analyse both the thermal and non-thermal emission processes that would be associated. Important to low-frequency radio observations is the synchrotron emission for which a formula is provided in Equation 32 of \cite{Nims2015ObservationalNuclei}: 

$$ \nu L_\nu = 10^{-5} \xi_{-2} L_{\mathrm{AGN}} \left(\frac{L_{kin}}{0.05 L_{\mathrm{AGN}}}\right) $$

Here the $\xi_{-2}$ parameter represents the percentage of the shock kinetic energy which is converted into relativistic electrons (i.e. if we assume 5\% of the energy goes into relativistic electrons, $\xi_{-2} = 5$). 

We attempt to test whether this model can account for all of the observed emission given the large number of quasars we have that are both radio-detected and have an estimate for their overall luminosity. We are also claiming that the \CIV distance traces the influence of winds, hence we can test whether the wind-dominated systems adhere to the \cite{Nims2015ObservationalNuclei} model more than the whole quasar population. 

To construct the predicted wind shock radio-luminosity at 144MHz from the \cite{Nims2015ObservationalNuclei} model we need to obtain the overall AGN luminosity, $L_{\mathrm{AGN}}$, and then pick appropriate values for the fraction of AGN energy in the wind and the fraction of wind energy that goes into relativistic electrons, $\xi_{-2}$. To obtain AGN luminosities we convert our interpolated six-micron luminosities from WISE and use the bolometric corrections from \cite{Richards2006}. We apply the same correction to all quasars, $L_{\mathrm{AGN}} = 8 \times L_{6\mu m}$, and then assume that $L_{kin} = 0.05 L_{\mathrm{AGN}}$.

We compute models for the expected wind shock radio luminosities at 144MHz for all of our quasars with a \CIV distance above 0.25, which should increase the prevalence of winds compared to jets. We create three models for the wind shock luminosity as follows: 

\begin{enumerate}
    
    \item Fixed conversion rate to relativistic electrons of $\xi_{-2} = 5$

    \item Varying the conversion rate $\xi_{-2}$ to minimise the Kolmogorov-Smirnov (K-S) test statistic between the observed radio luminosities and the model

    \item Giving the conversion rate a linear dependence on the AGN luminosity such that $\xi_{-2} = \frac{L_{\mathrm{AGN}}\alpha}{\max(L_{\mathrm{AGN}})} $ and fitting for $\alpha$ to minimise the K-S test statistic

\end{enumerate}

\begin{figure}
    \centering
    \includegraphics[width = 0.95\linewidth]{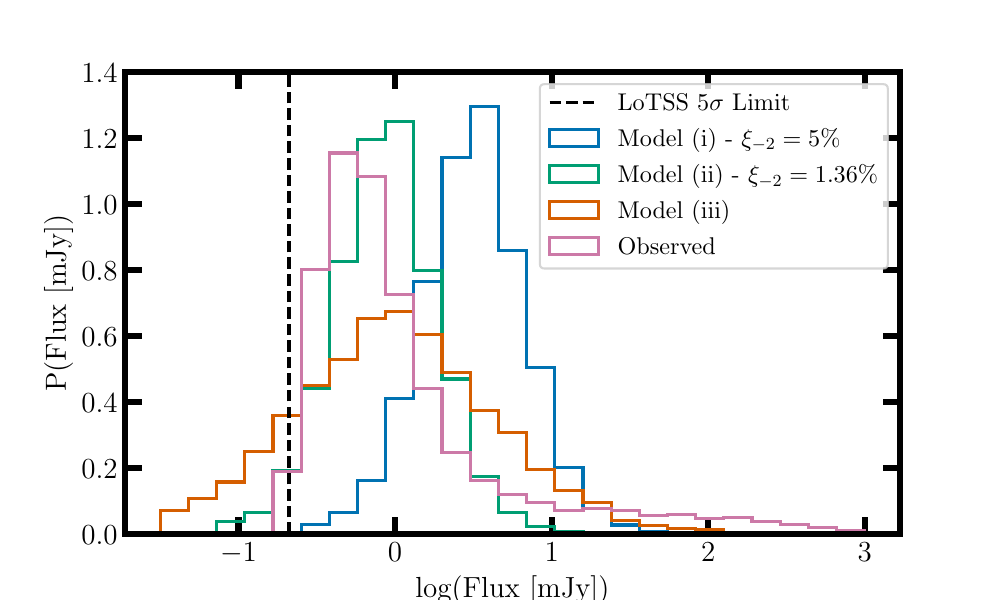}
    \caption{Three models of the wind shock radio flux as predicted by the prescription of \protect\cite{Nims2015ObservationalNuclei} using the AGN luminosity of radio-detected quasars. The observed radio flux is shown in pink, a constant 5\% of shock energy in relativistic electrons in blue, model~(i), a best fit constant conversion of 1.36\% in green, model~(ii), and a best fit functional form with a linear dependence on the AGN luminosity in red, model~(iii).}
    \label{fig:nims_model}
\end{figure}

The results of these different models along with the observed radio flux distribution are shown in \autoref{fig:nims_model}. We find that a similar assumption to the AGN to kinetic luminosity conversion of 5\%, model~(i), largely over-predicts the observed flux for our quasar sample. The value for this conversion factor that minimises the difference between the observed and model distributions is a more modest 1.36\%, model~(ii). The fit here is quite good, especially between the LoTSS flux limit and 10mJy. The fact that the shape is similar is not so surprising since the six-micron luminosity is roughly normal distribution and we are mostly shifting this distribution with some dependence on the redshift to obtain the fluxes. 

These models perform poorly for higher flux sources, which are a small fraction of the overall population, which suggests that wind shocks cannot explain all of the sources. We introduced the third model, model (iii), as a way to allow for potentially higher flux sources. This model includes a linear dependence of the efficiency on the AGN luminosity. Physically this would mean that higher luminosity AGN can convert more of their outflow energy to relativistic electrons. We do not have a specific assumption for how this would work but one example could be that a more powerful AGN creates a wider covering angle for outflows which allows for a larger shock front and more relativistic electrons. This model allows for the distribution to spread out in both directions. We do not recover the correct shape for the distribution and still cannot obtain the highest flux sources with just a linear connection between AGN luminosity and efficiency. Presumably, jets also contribute in some fraction of the sources so including some model for the jet flux could be an alternative way to fit the higher flux sources.

\section{Conclusions}

We have studied a sample of SDSS quasars within the LoTSS footprint. We have cut to a redshift range of $1.7<z<2.5$ to allow for BALQSO and quasar colour classification. We use spectral reconstructions as described in \cite{Rankine2020BALSample} to de-noise the spectra, improve the redshift estimates and characterise the \CIV emission more accurately. We look at trends for different quasar populations as a function of their \CIV distance, a tracer of the accretion rate. We focus on the radio properties of these sources as the origin of radio emission for most quasars is an open question. 

Our main findings are:

\begin{enumerate}
    \item Radio-detection fraction has a strong correlation with \CIV distance for all populations. Radio-detection fraction decreases with increasing \CIV distance up to 0.5 and then begins to increase. In the lowest \CIV distance bin the difference in radio properties for all groups narrows but this is also where we have the most uncertainty due to lack of sources. This narrowing could potentially be linked to the work of \cite{Temple2023TestingQuasars} where trends with accretion rate and black hole mass become apparent above some threshold;

    \item Radio luminosity does not change significantly with increasing \CIV distance while the radio-loudness of the populations decreases with increasing \CIV distance. This shows that g-band luminosity is increasing due to the changing shape of the SED of quasars across the \CIV distance space. This is further evidence that radio-loudness is not a suitable parameter to split radio populations without controlling for at least accretion rate.

    \item The colour of quasars is linked to their radio-detection fraction independent of \CIV distance and whether a BAL wind is present. The most reddened quasars have slightly higher radio luminosities than blue quasars when detected. The difference between moderately reddened and the most reddened quasars is less significant for BALQSOs;

    \item BALQSOs have a marginally higher radio-detection fraction than non-BALQSOs but overall the difference in detection fraction between these two populations is largely explained by their colour. This is further evidence that the BAL wind itself is connected to dust;

    \item The \cite{Nims2015ObservationalNuclei} wind shock model works well in modelling the observed radio-flux for radio-detected quasars. This is mainly just an appropriate scaling of the distribution of AGN luminosities, but the best fit relativistic electron coupling of 1.36\% is quite modest, giving wind shocks the potential to account for the radio emission of many sources;
    
\end{enumerate}

Are BALQSOs and red quasars distinct from the general population when it comes to their accretion and radio properties? For BALQSOs, their radio properties appear to be largely explained by their relation to optical colour. Although they have a preference for a particular region of the \CIV space, it is still true that BALQSOs exist with a wide range of \CIV EWs and blueshifts. It therefore appears that some combination of accretion rate, dust and orientation with respect to the observer is the origin of BALQSO winds. Red quasars have a similar connection between radio-detection fraction and accretion as blue quasars but then show different radio properties compared to the BALQSOs/non-BALQSOs. Their radio-detection enhancement is observable across \CIV space while detection fraction differences between red non-BALQSOs and red BALQSOs are narrower, indicating that the emission mechanism linked to colour could dominate over the BALQSO contribution.

Two key steps are needed to advance the studies of these populations with respect to their radio emission; one is observational and the other is theoretical. First, we need a statistically significant population of these populations to be studied at sub-galaxy scale resolution in the radio across the \CIV distance/accretion rate and optical colour space. This population will allow us to separate star formation and AGN components of the radio emission and then isolate the behaviour of the AGN component with an accretion state. Therefore, we should be able to distinguish between jets and wind shocks, a current limitation of our work. Secondly, we need more detailed theory and/or simulations of the behaviour and radiative transfer of wind shocks, which currently has the freedom to fit the entire population with the right energetic scaling, and the interaction of both wind shocks and jets with dust. Timescales are still a great uncertainty in this field and new simulations could be vital in constraining likely timescales for the presence of a BAL wind and the overlap with the radio emission it may be linked to or produce itself.

\section*{Acknowledgements}

JP acknowledges support for their PhD studentship from grants ST/T506047/1 and ST/V506643/1.

NLT and LKM acknowledge support from the Medical Research Council MR/T042842/1.

DMA acknowledges support from the STFC consolidated grant ST/T000244/1 and ST/X001075/1.

VAF acknowledges funding from a United Kingdom Research and Innovation grant MR/V022830/1.

PNB is grateful for support from the UK STFC via grant ST/V000594/1.

ALR acknowledges support from UKRI grant code MR/T020989/1.

IP acknowledges support from INAF under the Large Grant 2022 funding scheme (project “MeerKAT and LOFAR Team up: a Unique Radio Window on Galaxy/AGN co-Evolution”)

Funding for the Sloan Digital Sky Survey IV has been provided by the Alfred P. Sloan Foundation, the U.S. Department of Energy Office of Science, and the Participating Institutions. 

SDSS-IV acknowledges support and resources from the Center for High Performance Computing at the University of Utah. The SDSS website is \url{www.sdss.org}.

SDSS-IV is managed by the  Astrophysical Research Consortium for the Participating Institutions of the SDSS Collaboration including the Brazilian Participation Group, the Carnegie Institution for Science, Carnegie Mellon University, Center for Astrophysics | Harvard \& Smithsonian, the Chilean Participation Group, the French Participation Group, Instituto de Astrof\'isica de Canarias, The Johns Hopkins University, Kavli Institute for the Physics and Mathematics of the Universe (IPMU) / University of Tokyo, the Korean Participation Group, Lawrence Berkeley National Laboratory, Leibniz Institut f\"ur Astrophysik Potsdam (AIP),  Max-Planck-Institut f\"ur Astronomie (MPIA Heidelberg), Max-Planck-Institut f\"ur Astrophysik (MPA Garching), Max-Planck-Institut f\"ur Extraterrestrische Physik (MPE), National Astronomical Observatories of China, New Mexico State University, New York University, University of Notre Dame, Observat\'ario Nacional / MCTI, The Ohio State University, Pennsylvania State University, Shanghai Astronomical Observatory, United Kingdom Participation Group, Universidad Nacional Aut\'onoma de M\'exico, University of Arizona, University of Colorado Boulder, University of Oxford, University of Portsmouth, University of Utah, University of Virginia, University of Washington, University of Wisconsin, Vanderbilt University, and Yale University.

LOFAR data products were provided by the LOFAR Surveys Key Science project (LSKSP; \url{https://lofar-surveys.org/}) and were derived from observations with the International LOFAR Telescope (ILT). LOFAR \citep{vanHaarlem2013LOFAR:ARray} is the Low Frequency Array designed and constructed by ASTRON. It has observing, data processing, and data storage facilities in several countries, which are owned by various parties (each with their own funding sources), and which are collectively operated by the ILT foundation under a joint scientific policy. The efforts of the LSKSP have benefited from funding from the European Research Council, NOVA, NWO, CNRS-INSU, the SURF Co-operative, the UK Science and Technology Funding Council and the Jülich Supercomputing Centre.

The Jülich LOFAR Long Term Archive and the German LOFAR network are both coordinated and operated by the Jülich Supercomputing Centre (JSC), and computing resources on the supercomputer JUWELS at JSC were provided by the Gauss Centre for Supercomputing e.V. (grant CHTB00) through the John von Neumann Institute for Computing (NIC).

This research made use of Astropy,\footnote{https://www.astropy.org} a community-developed core Python package for Astronomy \citep{AstropyCollaboration2013Astropy:Astronomy,AstropyCollaboration2018ThePackage}.

For the purpose of open access, the author has applied a Creative Commons Attribution (CC BY) licence to any Author Accepted Manuscript version arising from this submission.

\section*{Data Availability}

The LoTSS DR2 radio data and catalogue used in this work are described in \cite{Shimwell2022VizieR2022} and can be found at the following page - \url{https://lofar-surveys.org/dr2_release.html}

The SDSS DR14 quasar catalogue used in this work is described in \cite{Paris2018TheRelease} is located at the following URL  - \url{https://data.sdss.org/sas/dr14/eboss/qso/DR14Q/DR14Q_v4_4.fits}

The SDSS spectra used as a basis for the ICA reconstructions can be downloaded through various means described on this page - \url{https://www.sdss.org/dr17/spectro/}.


\bibliographystyle{mnras.bst}
\bibliography{references_arxiv.bib}

\bsp	
\label{lastpage}
\end{document}